\documentclass[a4paper,11pt]{article}
\usepackage{jheppub}
\usepackage{amsmath}
\usepackage{amsfonts}
\usepackage{amssymb}
\usepackage{textcomp}
\usepackage{graphicx}
\usepackage{bm}
\usepackage{color}
\usepackage{verbatim}
\usepackage{subfig}
\usepackage{graphicx,color}
\usepackage{epsfig}

\usepackage{amsfonts,amsmath,amssymb}

\newcommand{\Pc}{{\cal P}}
\newcommand{\source}{\Sigma}

\begin{document}

\title{A simple model of momentum relaxation in Lifshitz holography}
 
\author[a]{Tom\'as Andrade}
 
\affiliation[a]{Rudolf Peierls Centre for Theoretical Physics \\ University of Oxford, 1 Keble Road, Oxford OX1 3NP, UK} 
  
\emailAdd{tomas.andrade@physics.ox.ac.uk}

\abstract{We expand the holographic studies of momentum relaxation to include non-relativistic scaling symmetries
in the ultraviolet. We do so by constructing black branes with Lifshitz asymptotics dressed with axions which explicitly 
depend on the boundary directions. Such configurations arise as analytic solutions of the Einstein-Proca theory coupled to massless
scalar fields in arbitrary dimensions. Studying linear perturbations on these backgrounds, we conclude that there is a 
dual Ward identity which accounts for the dissipation of momentum in the system. In addition, we numerically compute the 
frequency dependent thermal conductivity of the branes and verify that its DC limit is finite.}

\maketitle

\section{Introduction}

In the last few years the study of momentum relaxation in holography has received a considerable amount of attention. 
The main reason is that the gauge/gravity duality \cite{Maldacena:1997re, Witten:1998qj, Gubser:1998bc} provides with a powerful tool to describe transport properties in strongly coupled systems in which translational invariance is broken, hence coming closer to a more realistic description of condensed matter
systems. 
To achieve this, we must construct gravitational solutions which break translational symmetry along the boundary directions, 
which can be done by imposing boundary conditions that correspond  to turning on spatially dependent sources on the 
dual field theory.  
Generically, this entails solving a complicated system of non-linear, coupled, partial differential equations (PDE's).

For the case of periodic sources, this has been successfully carried out in 
\cite{Horowitz:2012ky} (see also \cite{Horowitz:2012gs, Donos:2014yya, Rangamani:2015hka, Chesler:2013qla, Ling:2013nxa}) 
with the expected results for the small frequency transport: the zero-frequency delta-function contribution in the conductivity, 
encountered when translational symmetry is preserved, is resolved into a Drude peak. 
The mathematical problem can be largely simplified by considering scenarios in which a global symmetry is present.
Then, one can break the translations precisely along this global symmetry direction, which implies that the 
resulting equations of motion are only ordinary differential equations (ODE's), since the explicit dependence on the 
boundary coordinates drops out \cite{Donos:2012js, Donos:2013eha, Andrade:2013gsa}, see \cite{Mateos:2011ix}
for an earlier example\footnote{An alternative holographic model for momentum relaxation was proposed in \cite{Vegh:2013sk}
which considered breaking diffeomorphism invariance by adding a mass to the graviton in four spacetime dimensions. 
In this approach the equations of motion for the background geometries are also ODE's. The theory of massive gravity was 
further studied in the holographic context in e.g. \cite{Blake:2013owa, Baggioli:2014roa, Baggioli:2015gsa, Alberte:2015isw}.}. Because of this, these configurations are usually referred to as ``homogeneous holographic lattices."
The model considered in \cite{Andrade:2013gsa} is of particular simplicity, consisting of Einstein-Maxwell theory coupled to a set of massless scalar fields which 
are taken to have linear dependence on the boundary coordinates and are thus termed {\it axions}. 
Homogeneity of the bulk stress tensor is ensured by the masslessness of the scalar fields, and they can moreover be  
arranged in such a way that an analytic solution with finite temperature and chemical 
potential can be found in arbitrary dimensions\footnote{This solution had previously been found in \cite{Bardoux:2012aw} although no holographic 
applications were considered there.}.
Interestingly, despite their mathematical simplicity the homogeneous lattices are able to capture the desired low energy physics that 
result from momentum relaxation and moreover display interesting phases including insulators, coherent and 
incoherent metals, superconductors, among others \cite{Gouteraux:2014hca, Donos:2014oha, Davison:2014lua, Davison:2015bea, Davison:2015taa, Taylor:2014tka, Erdmenger:2015qqa, Andrade:2014xca, Kim:2015dna, Ling:2014laa, Ling:2014saa, Ling:2015dma, 
Donos:2012wi, Donos:2013gda, Donos:2011ff, Donos:2013woa,Ammon:2016szz}\footnote{It appears, however, that these homogeneous 
lattices are not able to incorporate the physics of commensurability which one would expect of structures with a characteristic 
momentum scale \cite{Andrade:2015iyf}.}. 

So far, the focus of the study of momentum relaxation in holography has been placed on relativistic setups, i.e. the black hole geometries 
under scrutiny are asymptotically anti-de Sitter (AdS) so their dual field theories possess a scaling symmetry that treats time and space on the same footing. 
Since global symmetries need to match on each side of the duality, the first step in attempting to describe field theories 
endowed with other symmetries is to consider gravitational solutions which realize the algebra 
of interest as their asymptotic symmetry group. 
A thoroughly studied example is the $(d+1)$-dimensional Lifshitz metric \cite{Kachru:2008yh}, 
\begin{equation}\label{Lif metric}
 	ds^2 = \frac{dr^2}{r^2} - r^{2z} dt^2 + r^2 \delta_{a b} dx^a dx^b.
 \end{equation} 
Here $r$ is the holographic coordinate, chosen such that $r \to \infty$ describes the boundary of space-time, 
in which we introduce a time coordinate $t$ and $(d-1)$ spatial coordinates $x^a$.  
If the dynamical exponent $z$ is different from $1$, this metric is invariant under the following anisotropic scale transformation
\begin{equation}
	\vec x \to \lambda \vec x, \qquad t \to \lambda^z t , \qquad  r \to \lambda^{-1} r. 
\end{equation}
The holographic study of solutions that approach \eqref{Lif metric} at the boundary was suggested by \cite{Kachru:2008yh} and 
goes under the name of non-relativistic holography, or, more specifically, Lifshitz holography. See \cite{Taylor:2015glc} for 
a recent review\footnote{
It is also possible to consider solutions that possess other non-relativistic symmetries as part their asymptotic 
symmetry group. An example that has received much attention is the Schr\"odinger spacetime, the holographic studies 
of which were put forward in \cite{Son:2008ye, Balasubramanian:2008dm}. Here we will mostly be concerned with 
Lifshitz and only comment briefly on the Schr\"odinger case in section \ref{sec:conclusions}.}.

To facilitate the application of holographic techniques, it is convenient to think of \eqref{Lif metric} as the 
solution of a particular bulk theory. A popular bottom-up (i.e. non-stringy) model is the Einstein-Proca theory, 
in which the Lifshitz geometry is supported by a non-trivial massive vector \cite{Taylor:2008tg}\footnote{Here 
we adopt the point of view of  \cite{Kachru:2008yh} and consider relativistic gravitational field theories in the bulk 
which admit solutions with non-relativistic asymptotics. Another interesting approach is to take the gravitational bulk 
theories to also be non-relativistic \cite{Christensen:2013lma, Christensen:2013rfa, Hartong:2014pma, 
Hartong:2014oma, Hartong:2015wxa, Hartong:2015zia}.}.
Finite temperature generalizations of \eqref{Lif metric} have not been found analytically in the Einstein-Proca theory 
and so far these have been obtained resorting to other somewhat unusual modifications of Einstein gravity.
These include Einstein-Proca coupled to a scalar field without a kinetic term \cite{Balasubramanian:2009rx}, 
gravity with higher curvature interactions \cite{AyonBeato:2009nh, Cai:2009ac, AyonBeato:2010tm, Dehghani:2010kd, 
Matulich:2011ct, Brenna:2011gp, Gonzalez:2011nz, Dehghani:2011hf, Liu:2012yd, Lu:2012xu, Ghanaatian:2014bpa},
Brans-Dicke theory \cite{Maeda:2011jj}, 
and higher spin gravity \cite{Gutperle:2013oxa, Beccaria:2015iwa}. Additionally, Lifshitz black branes have been constructed in 
the presence of dilatonic couplings with running scalar fields, see e.g. \cite{Taylor:2008tg, Tarrio:2011de}.

In order to extract the physics of the dual theory, we need to identify the independent pieces of the 
asymptotic UV data and associate them with sources and vevs of the dual field theory operators, 
i.e. solve the {\it dictionary} problem.
Regarding this, the main difference of non-relativistic holography with its relativistic counterpart is the fact that, 
due to the anisotropic scaling, the former lacks a well-defined conformal boundary metric, as can be explicitly seen in 
\eqref{Lif metric}. An interesting resolution of this issue is to work in terms of the veilbien fields instead of the metric, 
which allows for a clean separation of the temporal and spatial pieces of the geometry \cite{Ross:2009ar, Ross:2011gu}. 
This separation naturally implements the fact that, as expected in a non-relativistic field theory, the 
relativistic (symmetric) stress-tensor needs to be replaced by a stress tensor complex, in which, 
for example, momentum and energy fluxes enter differently. 
Adopting this approach, the dictionary problem for Lifshitz was worked out in \cite{Ross:2009ar} 
by studying linear perturbations around the pure Lifshitz background \eqref{Lif metric} embedded in 
the Einstein-Proca theory.
Their strategy can be summarized as follows: 
extending the usual relativistic AdS/CFT dictionary, sources of the stress tensor complex are associated with 
perturbations that shift the appropriate components of the boundary vielbien. Then, after constructing an appropriate action 
principle, vevs are obtained as the variations of the action with respect to the sources. Alternatively, as suggested in 
\cite{Papadimitriou:2010as}, vevs can be identified as the boundary quantities which are canonically conjugated to the sources
in the symplectic flux evaluated at the boundary, which circumvents the computation of the counterterms at least 
in a linear approximation. This procedure was carried out in \cite{Andrade:2013wsa} for the Lifshitz background 
and we will employ the same method here. The dictionary problem for Lifshitz was further elucidated and 
extended to hyperscaling geometries in \cite{Chemissany:2014xpa, Chemissany:2014xsa}.

In this paper we initiate the study of momentum relaxation in non-relativistic holography. The first step will 
be to construct black brane solutions which break translational invariance and asymptote to \eqref{Lif metric} near the boundary.
By extending the mechanism of \cite{Andrade:2013gsa} to non-relativistic setups, this can be easily done by 
adding axions to the Einstein-Proca model. 
Interestingly, in addition to breaking translational invariance 
this allows us to find an analytic finite temperature solution, absent in the theory when the scalars vanish. 
The second step will be to demonstrate that the so constructed black holes constitute in fact systems that violate momentum 
conservation.
To do so, we will show that the Ward identity that yields momentum conservation in the translationally invariant 
solution is modified in the presence of the axions in such a way that the violation of conservation is proportional 
to the product of the sources and vevs dual to the axions. In close parallel to the relativistic case, this Ward identity becomes 
non-trivial at the level of linear fluctuations.
As a compelling consistency check we will numerically compute the AC thermal conductivity, which is related to the 
two point function of the momentum operator by a Kubo formula, and explicitly verify that it approaches a finite DC value. 

This paper is organized as follows: in section \ref{sec:backgnd} we present the analytic solutions 
of interest and discuss some of their properties. We analyze the linear perturbations that capture the physics of the dual 
momentum operator in section \ref{sec:momentum relax}, where we derive the Ward identity that shows that momentum is 
not conserved, and compute the thermal conductivity. We conclude and discuss some open issues in 
section \ref{sec:conclusions}. Finally, we provide two appendices in which we collect useful results regarding the linear fluctuations. 
  
\section{Analytic Lifshitz black branes with scalar sources}
\label{sec:backgnd}

As mentioned above, the Lifshitz metric \eqref{Lif metric} can be obtained as an analytic solution in the 
Einstein-Proca theory with negative cosmological constant. In order to incorporate momentum relaxation in the system 
it is natural to add massless axions to the setup, in the way suggested in \cite{Andrade:2013gsa}. This leads us to consider the 
following action in $(d+1)$ dimensions
\begin{equation}\label{action}
	I = \int d^{d+1} x \sqrt{- g} \left[ R - 2 \Lambda - \frac{1}{4} F^2 - \frac{m^2}{2} A^2 - 
	\frac{1}{2} \sum_{I=1}^{d-1} (\partial \psi_I)^2 \right], 
\end{equation}
\noindent where $R$ is the Ricci scalar, $\Lambda$ is the cosmological constant, $A$ is a massive vector field
with mass $m$ and field strength $F = dA$, and $\psi_I$ is a collection of $(d-1)$ massless scalar fields collectively 
referred to as axions. 
Throughout this paper we will be concerned with configurations of the form 
\begin{equation}\label{ds2 ansatz}
	ds^2 = \frac{dr^2}{h(r)} - f(r) dt^2 + r^2 \delta_{a b} d x^a dx^b ,
\end{equation}
\begin{equation}\label{A and psi} 
	A = a(r) dt , \qquad \psi_I = \alpha \delta_{I a} x^a , 
\end{equation}
\noindent where $\alpha$ is a dimension one constant that controls the amount of  translational symmetry breaking. 
Within this ansatz, we can readily check that we obtain an analytic solution of the equations of motion by letting 
\begin{equation}\label{solnf}
	f(r) = r^4 \left (1 - \frac{r_0^2}{r^2} \right), \qquad h(r) = r^2\left(1 - \frac{r_0^2}{r^2} \right), 
	\qquad  a(r) = r^2\left(1 - \frac{r_0^2}{r^2} \right),
\end{equation}
\noindent with
\begin{equation}\label{soln const}
	m^2 = 2 (d-1), \qquad \Lambda = - \frac{1}{2} (d^2+1), \qquad r_0 = \frac{\alpha}{\sqrt{2(d-1)}} .
\end{equation}
Here we have set the overall curvature length scale to one, as we shall do henceforth. 
The solution \eqref{ds2 ansatz}, \eqref{A and psi}, \eqref{solnf} asymptotes to the Lifshitz metric \eqref{Lif metric} with $z=2$, and possesses 
a regular black hole horizon at $r = r_0$ with temperature 
\begin{equation}
	T = \frac{1}{4 \pi} \sqrt{ \frac{h(r)}{f(r)}} f'(r) \bigg|_{r = r_0} = \frac{1}{4\pi} \frac{\alpha^2}{(d-1)} . 
\end{equation}
Note that the temperature has dimension 2 due to the Lifshitz scaling. 
The metric \eqref{ds2 ansatz} is homogeneous and isotropic in the spatial coordinates, although the full solution 
is not invariant under spatial translations due to the presence of the scalar fields \eqref{A and psi}, just as in the asymptotically 
AdS case studied in \cite{Andrade:2013gsa}. It is worth noting that, as opposed to the $z=1$ solutions of the Einstein-axion model, 
no logarithmic branches appear when we take $d=2$. 

By solving the Klein-Gordon equation for a scalar of mass  $m_\psi$ in the spacetime \eqref{Lif metric}, 
we find that the two independent falloffs correspond to scaling dimensions \cite{Kachru:2008yh}
\begin{equation}
	\Delta_\pm = \frac{1}{2}(z + d-1) \pm \sqrt{ m_\psi^2 + \frac{1}{4} (z + d-1)^2 }, 
\end{equation}
The standard AdS/CFT dictionary instructs us to identify the leading (subleading) coefficient with the source (vev) 
of a dual scalar operator. In the case at hand we have $m_\psi^2 = 0$, $z = 2$, so that 
$\Delta = 0, (d+1)$. Hence, we can interpret the axions $\psi \sim \alpha x$ as turning on spatially-dependent
scalar sources for $(d-1)$ dual scalar operators ${\cal O}_I$ of dimension $(d+1)$. 

Note that, because of the relations \eqref{soln const}, $\alpha$ is the only free parameter in the solution, 
which can be moreover scaled away a coordinate transformation.
In this sense, the geometry above is the analogue of the relativistic solution of \cite{Andrade:2013gsa} with the energy and
chemical potential set to zero. Although we have not computed the charges of these black branes in detail, we can 
argue on dimensional grounds that they have zero energy. First, note that in theories with Lifshitz scaling the energy density 
has dimensions $(z + d-1)$. There are no integration constants with 
such dimension in the functions \eqref{solnf}, which suggests that the black branes \eqref{ds2 ansatz}, \eqref{A and psi}, \eqref{solnf}
indeed have zero energy. Similar solutions have been obtained analytically in \cite{Mann:2009yx}, which 
considers black holes with hyperbolic horizons instead of the axionic coupling.
%

As mentioned in the introduction, the black brane geometry \eqref{ds2 ansatz}, \eqref{solnf}
had previously been obtained in arguably non-standard theories of gravity, some of which 
are expected to be pathological.   
Remarkably, besides breaking translational invariance, the presence of the axions allows for 
an analytic finite temperature generalization of \eqref{Lif metric}, so that the configuration 
\eqref{ds2 ansatz}, \eqref{A and psi}, \eqref{solnf}  arises naturally in a theory which appears 
to be well-behaved and has a transparent interpretation in terms of a dual theory. 
We expect the dual field content to simply be that of the Einstein-Proca theory --  the stress tensor
complex along with a scalar operator \cite{Ross:2009ar} -- in addition to a set of scalar operators
of dimension $(d+1)$ introduced by the axions. We will verify that this is the case in the sector of perturbations
dual to the momentum operator. 

In $d=3$, it is possible to find another analytic solution which has Lifshitz asymptotics with $z=4$. 
This configuration takes the form \eqref{ds2 ansatz}, \eqref{A and psi}, with 
\begin{equation}
   f(r) = r^6 h(r) , \qquad a(r) = r^2 h(r), \qquad h(r) = \frac{3}{2} (r^2 - r_0^2) \left(1 + \frac{r_0^2}{3 r^2} \right) , 
\end{equation} 
\noindent and 
\begin{equation}\label{soln const 2}
	m^2 = 12 , \qquad \Lambda = - 18 , \qquad r_0 = \frac{\alpha}{\sqrt{20}} .
\end{equation}
The temperature is given by 
\begin{equation}
	T = \frac{\alpha^4}{400 \pi} .
\end{equation}

\noindent Just like in the family of $z=2$ solutions above, there are no integration constants of 
dimension $(z + d-1) = 6$, so we expect this $z=4$ brane to also have zero energy.
While finding more general solutions in an analytic way has escaped us, we expect their numerical construction 
to be relatively straightforward, which we propose as a future direction of this work. 

The fact that the solutions presented here break translational invariance strongly suggests that they constitute holographic 
systems in which momentum is not conserved. We will explicitly show that this is indeed the case in the following section.

\section{Momentum relaxation}
\label{sec:momentum relax}

This section is devoted to demonstrate that the Lifshitz black holes of section \ref{sec:backgnd} 
are in fact geometries that implement momentum relaxation. In order to do so, we study the linear fluctuations that control 
the dynamics of the momentum density operator $\Pc_a$.
More precisely, we will obtain the Ward identity that shows the non-conservation of momentum at the linearized level
in addition to explicitly verifying that the zero frequency limit of the thermal conductivity is finite. 
For concreteness, we will focus on the case of $z=2$, $d=3$. We believe that our procedure below should be applicable
for other values of the parameters with minor modifications. 

\subsection{Ward identity}
\label{sec:ward}

As argued in \cite{Ross:2009ar}, the fluctuations of the $g_{tx}$ component of the metric
encode the dynamics of $\Pc_x$, so we concentrate on this type of perturbations\footnote{By isotropy 
of the background we are free to choose any component of the momentum to perform the analysis. 
Throughout this section we refer to this direction as $x$, without extra indices, for notational convenience.}.
We obtain a consistent set of linearized equations about the background  \eqref{ds2 ansatz}, \eqref{A and psi}, \eqref{solnf}  
by considering
\begin{equation}\label{pert Px}
	\delta g_{tx} = e^{- i \omega t} r^2 h_{tx}(r) , \qquad \delta A_x = e^{- i \omega t} a_x(r), \qquad \delta \psi_1 =  e^{- i \omega t} s(r).
\end{equation}
\noindent We have included the prefactor $r^2$ in $\delta g_{tx}$ to simplify the expressions below.
The equations of motion that govern the dynamics of this set of perturbations are
\begin{align}\label{eqlin1}
	s'' + \left( \frac{3}{r} + \frac{h'}{h} \right)s' + \frac{\omega^2}{r^2 h^2} s - \frac{i \alpha \omega}{r^2 h^2} h_{tx} &= 0, \\
	a_x'' + \left( \frac{1}{r} + \frac{h'}{h} \right) a_x' + \left( \frac{\omega^2}{r^2 h^2} - \frac{4}{h} \right) a_x + \frac{a'}{h} h_{tx}' &= 0 ,\\
\label{eqlin3}
\alpha s - \frac{i \omega}{h} h_{tx}' - \frac{i \omega a'}{r^2 h} a_x &= 0.
\end{align}
In order to simplify the equations above we have used that $a = r^{-2} f = h$, which directly follows from \eqref{solnf}. 
The UV asymptotics of the linearized fields are given by
\begin{align}
\label{uv series1}
	h_{tx} &= r^2\left( h_{tx}^{(0)} + \frac{h_{tx}^{(1)}}{r^2} +  \frac{h_{tx}^{(2)}}{r^4} +  
	\frac{\log r}{r^4} \tilde h_{tx}^{(2)}  + \frac{h_{tx}^{(3)}}{r^6} + \ldots \right),  \\ 
 a_x &= r^2\left( a_{tx}^{(0)} + \frac{a_{x}^{(1)}}{r^2} +  \frac{a_{x}^{(2)}}{r^4} +  
	\frac{\log r}{r^4} \tilde a_{x}^{(2)}  + \frac{a_{x}^{(3)}}{r^6} + \ldots \right) , \\
\label{uv series3}
s &= s^{(0)} + \frac{s^{(1)}}{r^2} +  \frac{s^{(2)}}{r^4} +  
	\frac{\log r}{r^4} \tilde s^{(2)}  + \frac{s^{(3)}}{r^6} + \ldots \, .
\end{align}
The logarithms appear because we are working with integer $z$ and they signal the presence of
a scale anomaly \cite{Baggio:2011ha, Chemissany:2012du}. 
The independent pieces of the UV asymptotic data can be taken to be $h_{tx}^{(0)}$, $h_{tx}^{(1)}$, 
$h_{tx}^{(2)}$, $h_{tx}^{(3)}$ and $s^{(0)}$, with the remaining coefficients being fixed by the asymptotic equations of motion. 
The explicit expression for the first few orders are given in appendix \ref{app:asympt}. 
It is worth noting that the ansatz \eqref{pert Px} admits a residual symmetry generated by the vector
$\xi = e^{- i \omega t} \xi^x_0 \partial_x$,  where $\xi^x_0$ is a constant. This gauge symmetry acts non-trivially on the coefficients as
$\delta s^{(0)} = \alpha \xi^x_0$, $\delta h_{tx}^{(1)} = - i \omega \xi^x_0$ which implies that the combination 
\begin{equation}\label{gauge inv source}
	\source^{(0)}:=  h_{tx}^{(1)} + i \frac{\omega}{\alpha} s^{(0)}
\end{equation}
\noindent is gauge invariant. 

In order to identify the dual sources and vevs in terms of the coefficients of the UV expansion
we proceed as in \cite{Andrade:2013wsa}. First, by noting which fluctuations modify the different components of the
boundary vielbeins, we conclude that $h_{tx}^{(0)}$ and $h_{tx}^{(1)}$ are sources for the energy flux and 
momentum density, respectively. Moreover, as stated in section \ref{sec:backgnd}, we relate the leading branch of 
the scalar field $s^{(0)}$ to the source of a dual scalar operator. 
Following \cite{Andrade:2013wsa}, we henceforth set $h_{tx}^{(0)} $ to zero since it is the source for the energy 
flux which is an irrelevant operator. 

Having found the sources we proceed to read off the vevs by evaluating the symplectic flux at the boundary. 
The computation is rather technical so we leave the details for appendix \ref{app:flux}. 
The key result is that we can identify the vevs of the momentum operator ${\cal P}_x$ and 
the scalar operator ${\cal O}$, dual to the fluctuating axion, in terms of the coefficients of the boundary expansions as
\begin{align}\label{P and O1}
 	{\cal P}_x &= 2 a^{(2)}_x - 2 h_{tx}^{(2)} + \tilde h_{tx} ^{(2)} = - 4 h_{tx}^{(2)} - \frac{3 \alpha^2}{4} \source ^{(0)},   \\
 \label{P and O2}
 	{\cal O} &=  4 s_{(2)} - \tilde s^{(2)} = \frac{i \omega}{\alpha} \left(4 h_{tx}^{(2)} + \frac{3 \alpha^2}{4} \source ^{(0)} \right).
\end{align} 
It is worth commenting that, up to terms local in the sources, this matches our expectations from previous results in the literature. Firstly, 
from the free Klein-Gordon field calculation of \cite{Kachru:2008yh}, we expected ${\cal O} \sim s_{(2)}$, which indeed holds true. Moreover, 
\cite{Ross:2009ar, Andrade:2013wsa} found that ${\cal P}_x \sim h_{tx}^{(2)}$ which is also verified in our computation. The energy flux
is related to the coefficient $h_{tx}^{(3)}$ but the details of this part of the dictionary will not concern us here. 

Aimed with equations \eqref{P and O1} and \eqref{P and O2}, we can now verify that the following relation holds
\begin{equation}\label{Ward}
	\partial_t {\cal P}_x = \partial_x \psi^{(0)} {\cal O}, 
\end{equation}
\noindent which is the sought for Ward identity that accounts for momentum relaxation in the system. Here we have 
transformed back to position space noting that the background value for the source of the scalar, denoted by $\psi^{(0)}$,
satisfies $\partial_x \psi^{(0)} = \alpha$. Denoting the spatial stress tensor by $\pi^{a b}$, 
we expect the term $\partial_a \pi^{a} _x$, to appear on the left hand side of \eqref{Ward} when spatially dependent perturbations
are taken into account \cite{Ross:2009ar, Ross:2011gu}. Because holographic Ward identities are the result 
of diffeomorphism invariance of the on-shell action \cite{deHaro:2000vlm}, it is not surprising for 
\eqref{Ward} to adopt the same form as in the relativistic case discussed in \cite{Andrade:2013gsa} 
at zero chemical potential.

\subsection{Thermal conductivity}

\begin{figure}[ht]
\centering
 \includegraphics[width=1\linewidth]{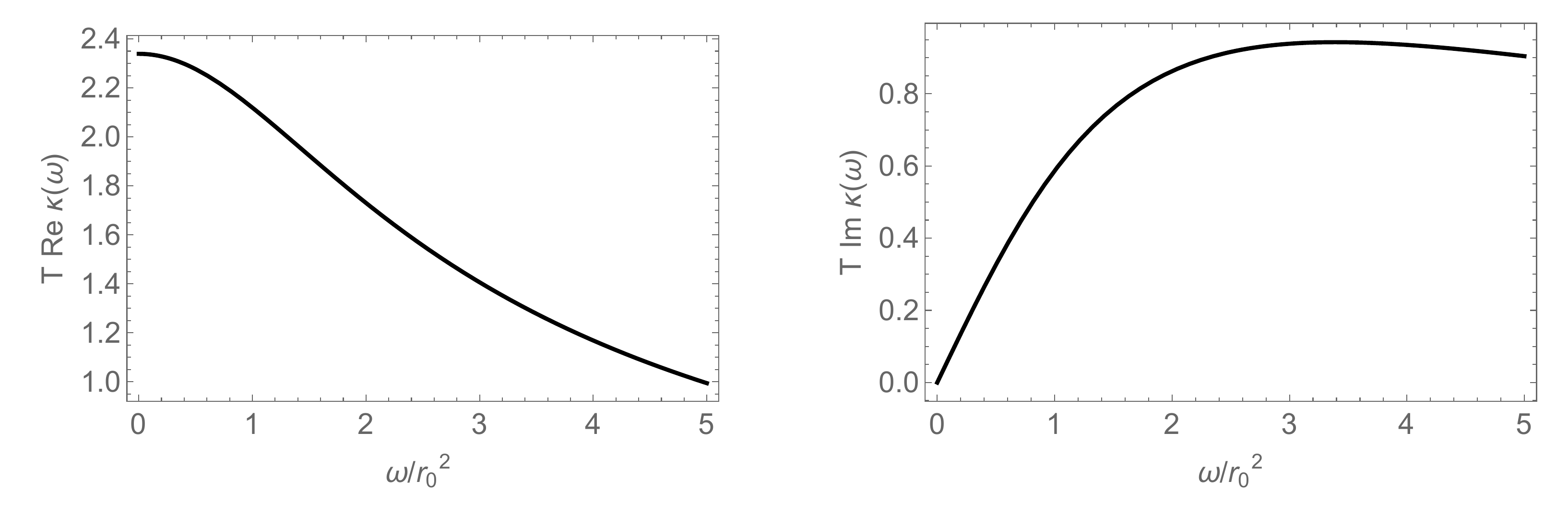}
  \caption{\label{kappa plots} Real (left) and imaginary (right) parts of the frequency dependent thermal conductivity. Quantities on both 
  axes are scale-invariant.}
\end{figure}

We will now numerically calculate the frequency-dependent thermal conductivity, $\kappa(\omega)$, for the spatially homogeneous perturbations described above. Given the two-point function of the 
momentum operator, which we denote in momentum space by $G(\omega)$, the thermal conductivity 
at zero chemical potential can be found via the Kubo formula
\begin{equation}\label{kappa kubo}
	\kappa (\omega) = \frac{i}{\omega T} [ G(\omega)- G(0) ], 
\end{equation}
\noindent which implies that $\kappa$ has a dimension $-2$.
From the analysis of section \ref{sec:ward}, it is clear how to calculate $G(\omega)$: in the linear approximation, this is simply the ratio
between the vev and source of the momentum operator which are known in terms of the boundary data. 
In order to obtain a gauge-invariant result, the source we need to consider
is in fact the combination \eqref{gauge inv source}. Hence, it follows from \eqref{P and O1} that the expression for the 
Green's function in terms of the coefficients of the asymptotic expansion is given by 
\begin{equation}\label{G}
	G(\omega) = -  \frac{4 h_{tx}^{(2)}}{\source ^{(0)}}.
\end{equation}
Here we have used the renormalization scheme freedom to drop local contributions to the one-point function  
\cite{deHaro:2000vlm}. 
We have explicitly checked that their inclusion does not affect the results obtained below. Expression \eqref{G} is to be evaluated
for solutions in which the other field theory source, $h^{(0)}_{tx}$, is set to zero. Moreover, we must impose ingoing boundary conditions
at the black hole horizon to ensure that the holographic prescription gives the retarded correlator \cite{Son:2002sd}. 
With these conditions, the right hand side of \eqref{G} is uniquely determined once a value of $\omega$ is given\footnote{It 
is worth noting that $\alpha$ can be set to one by simple coordinate transformations, so the only free parameter in 
the numerical calculation is $\omega$.}.
We extract $G(\omega)$ by numerically solving the equations of motion \eqref{eqlin1}-\eqref{eqlin3} employing a shooting method, 
and then use \eqref{kappa kubo} to obtain the AC thermal conductivity. Our results are displayed in figure \ref{kappa plots}. 

As expected of a system which breaks translational invariance,
the thermal conductivity does not present a delta function at zero frequency, 
as can be seen from the fact that ${\rm Im} \, \kappa(\omega) \to 0$ as $\omega \to 0$\footnote{Recall that the 
Kramers-Kronig relation implies that a delta function contribution to ${\rm Re} \, \kappa(\omega) $ would manifest
itself as a pole in ${\rm Im} \, \kappa(\omega)$ \cite{Hartnoll:2009sz}.}. 
Instead, we find a finite peak in ${\rm Re} \, \kappa(\omega) $, which can be thought of as the resolution of the $\delta (\omega)$
present in the momentum-conserving system. 
For small $\alpha$, the peak of the thermal conductivity in the relativistic Einstein-axion model can be well 
approximated by the Drude formula
\begin{equation}\label{Drude}
	\kappa(\omega) = \frac{\kappa(0)}{1 - i \omega \tau},
\end{equation}
\noindent where $\tau$ is a characteristic time. As we increase $\alpha$, the peak broadens until we reach a 
regime in which the Drude approximation fails \cite{Kim:2014bza}, 
signalling a transition from a coherent to an incoherent phase\footnote{We refer the reader to \cite{Hartnoll:2014lpa} 
for a discussion on this terminology.}. 
Coherence at small $\alpha$ stems from the fact that momentum is a quasi-conserved quantity. In this case, using the 
memory matrix formalism in field theory,
it is possible to conclude that \eqref{Drude} must hold
\cite{2007PhRvB..76n4502H, Hartnoll:2012rj, 2013PhRvB..88l5107M}.
As pointed out in \cite{Davison:2014lua}, 
the (in)coherence of the system can be understood from the structure of the quasi-normal modes (QNM). In fact,
in the coherent regime transport is controlled by a single, purely-dissipative, long-lived excitation which as such gives rise to 
\eqref{Drude}. As $\alpha$ increases, this dissipative mode mixes with other excitations in the spectrum so transport is no 
longer coherent. 

Coming back to the non-relativistic case, we observe that the width of the peak in figure \ref{kappa plots} 
is large, which indicates that we are in an incoherent regime. 
Indeed, computing $G(\omega)$ in the complex plane, we do not find QNM
near the origin, so there is no evidence of isolated excitations controlling the heat transport
in the system\footnote{Similarly, zero energy solutions on the Einstein-axion model were found to transport heat 
incoherently \cite{Davison:2014lua}.}. Since $\alpha$ is the only parameter of the background solution, scale invariance implies that
all solutions with non-zero $\alpha$ are equivalent, so we cannot access the coherent regime in this analysis. 
Solutions with tunable $\alpha$ should be easy to find numerically, and they would help us 
search for a coherent/incoherent transition in this non-relativistic holographic system.

\section{Conclusions}
\label{sec:conclusions}

We have found a family of $z=2$ Lifshitz black branes in arbitrary dimensions as analytic solutions 
of the Einstein-Proca theory coupled to massless scalar fields which depend linearly on the 
boundary coordinates. 
By studying linear fluctuations on this background for $d=3$, we have concluded that there is a Ward identity
which explicitly shows that momentum relaxes as a consequence of the breaking of translational invariance. 
Moreover, also in $d=3$, we have computed the frequency dependent thermal conductivity numerically and verified that its DC limit is finite. 
The broadness of the zero frequency peak in the thermal conductivity indicates that transport is incoherent in this background.
We have also found an analytic finite temperature solution with $z=4$ and $d=3$. We expect this configuration and 
the higher dimensional $z=2$ ones to behave similarly to the $z=2$, $d=3$ case studied in detail here.

In order to obtain analytic results, we have restricted ourselves to uniparametric backgrounds 
characterized by $\alpha$ solely, which appear to be dual to zero energy configurations. By scale invariance,
all solutions with non-zero $\alpha$ are equivalent, so, for a given dimension, these solutions are in fact 
one and the same. Despite the non-relativistic asymptotics of such configurations, their properties seem to closely 
resemble those of their relativistic analogue in the Einstein-axion theory. 
Finding more general black hole solutions with Lifshitz asymptotics and dressed with linear axions 
is a straightforward numerical task since it amounts to solving non-linear ODEs. Having more free parameters, 
these configurations would be relevant to elucidate the main differences between the relativistic and non-relativistic 
models of this type. We leave this interesting problem for future work.

As noted in \cite{Giacomini:2012hg}, the Klein Gordon equation for a field of mass squared $m_\psi^2$  can be solved 
analytically on the $z=2$ backgrounds of section \ref{sec:backgnd}. Using this, \cite{Giacomini:2012hg} concluded 
that these geometries are stable for Dirichlet boundary conditions -- which set to zero the leading branch of the field
near the boundary --   if $m_\psi^2 > 0$\footnote{Other choices of boundary conditions for scalar fields in Lifshitz were
considered in \cite{Keeler:2012mb, Andrade:2012xy}, which determined a range of masses in which the slow fall off of the field 
is allowed to fluctuate by normalizability. Somewhat surprisingly, even within this range of masses one finds unstable
modes of the form $\omega \sim i k^z$ \cite{Andrade:2012xy}. }. 
At least in the pure Lifshitz geometry, the case $m_\psi^2 < 0$ is also well defined for Dirichlet boundary conditions
if the analogue of the Breitenlohner-Freedman (BF) bound in AdS \cite{Breitenlohner:1982bm, Breitenlohner:1982jf} is 
respected \cite{Kachru:2008yh}, making a closer inspection of this case worthwhile. 
It turns out that for spacetime dimensions higher than four and negative squared masses above the Lifshitz BF bound,
the aforementioned $z=2$ black holes can be unstable against scalar perturbations which have small spatial 
momentum \cite{Andrade:2012xy}. 
Given this, it seems pertinent to investigate if the presence of fields with vanishing mass in the Einstein-Proca-axion 
model could lead to instabilities of the kind discussed above for high enough dimensions.

To our knowledge, our construction constitutes the first example of a non-relativistic holographic model 
of momentum relaxation. Wondering about the generality of the procedure employed here, it is 
natural to ask whether or not the coupling to the axions yields similar configurations in other 
non-relativistic setups.
For instance, it would be interesting to find a (perhaps finite temperature) analytic generalization of the Schr\"odinger spacetime 
which includes axions. 

\acknowledgments

It is a pleasure to thank Eloy Ay\'on-Beato, Mokhtar Hassa\"ine and Ben Withers for valuable discussions,
and especially Lorena Andrade for suggesting this research project. 
This work was supported by the European Research Council under the European Union's Seventh Framework Programme
(ERC Grant agreement 307955). 
The partial support of the Newton-Picarte Grant 20140053 and the hospitality of the organizers of the conference 
``Spacetimes with anisotropic scaling symmetry and holography", which took place in Vi\~na del Mar, Chile, on January 2016 
are also gratefully acknowledged.

\appendix

\section{Asymptotics of the linearized perturbations}
\label{app:asympt}

Here we gather the results for the boundary series expansions of the linear fluctuations studied in section \ref{sec:ward}.
Inserting the expansions \eqref{uv series1}-\eqref{uv series3} into \eqref{eqlin1}-\eqref{eqlin3} we find the following 
relations among the coefficients
\begin{align}
\label{UV coeff 1}
	a_x^{(0)} &=  - h_{tx}^{(0)} , \\
	a_x^{(1)} &= \frac{\alpha^2}{4} h_{tx}^{(0)},  \\
	s^{(1)} &=	 - \frac{i \omega}{4} h_{tx}^{(0)},   \\
	a_x^{(2)} &=  - \frac{1}{16} [ 16 h_{tx}^{(2)} + 4 \alpha ( \alpha h_{tx}^{(1)} + i \omega s^{(0)} ) + 
	 ( \alpha^4 + 4 \omega^2 ) h_{tx}^{(0)}  ] ,  \\
	s^{(2)} &= i \frac{\omega}{8 \alpha} [ 8 h_{tx}^{(2)} + \alpha  ( \alpha h_{tx}^{(1)} + i \omega s^{(0)} ) + \omega^2 h_{tx}^{(0)}   ] ,  \\
	\tilde h_{tx}^{(2)} &= - \frac{\alpha}{16} [  4  (\alpha h_{tx}^{(1)} + i \omega s^{(0)} ) + \alpha^3 h_{tx}^{(0)}  ],  \\
	\tilde a_{x}^{(2)} &=  \frac{\alpha}{16} [  4  (\alpha h_{tx}^{(1)} + i \omega s^{(0)} ) + \alpha^3 h_{tx}^{(0)}  ] ,  \\
	\tilde s^{(2)} &= - \frac{i \omega}{16} [  4  (\alpha h_{tx}^{(1)} + i \omega s^{(0)} ) + \alpha^3 h_{tx}^{(0)}  ]  .
\label{UV coeff -1}
\end{align}

\section{Symplectic flux and conjugate pairs}
\label{app:flux}

In this appendix we present the details of the  symplectic flux calculation quoted in section \ref{sec:ward}. 
Given two linearized solutions, $\delta_{\bf 1}$ and $\delta_{\bf 2}$\footnote{The quantities $\delta_{{\bf 1}, {\bf 2}}$ denote linearized configurations in which all fields fluctuate, i.e. $\delta_{\bf 1} = \{\delta_{\bf 1} g, \delta_{\bf 1} A, \delta_{\bf 1} \psi_I \}$ and likewise for $\delta_{\bf 2}$. Boldface subindices label linear fields in each linearized solution.}, the symplectic
current, $j^\mu$, can be constructed employing the algorithm detailed in \cite{Lee:1990nz}. 
The explicit expression for the Einstein-Proca theory can be found in \cite{Andrade:2013wsa}, here we only need to supplement it by adding 
a Klein-Gordon piece due to the presence of the scalars. The full result is then 
\begin{equation}
	j^\mu = j^\mu_g + j^\mu_A + j^\mu_\psi, 
\end{equation}
\noindent where $j^\mu_g$, $j^\mu_A$ and $j^\mu_\psi$ are the gravitational, vectorial and scalar contributions, respectively given by
\begin{align}
 	j^\mu_g &= \delta_{\bf 2} (\sqrt{- g} g^{\alpha \beta}) \delta_{\bf 1} \Gamma^\mu  _{\alpha \beta} - \delta_{\bf 2}(\sqrt{- g} g^{\mu \alpha}) \delta_{\bf 1} \Gamma^\beta_{\alpha \beta} - ({\bf 1} \leftrightarrow {\bf 2} ), \\
    j^\mu_A &= - \delta_{ \bf 2} (\sqrt{- g} F^{\mu \nu}) \delta_{\bf 1}  A_\nu  - ({\bf 1} \leftrightarrow {\bf 2} ) , \\
 j^\mu_\psi &= - \sum_{I=1}^{d-1}\delta_{\bf 2} (\sqrt{- g} \partial^\mu  \psi_I ) \delta_{\bf 1} \psi_I  - ({\bf 1} \leftrightarrow {\bf 2}).
\end{align} 
\noindent Note that we are using conventions in which the symplectic current is a vector density. We can readily check 
that the total symplectic current is closed on-shell, $\partial_\mu j^{\mu} = 0$. 

The boundary symplectic flux ${\cal F}$ is defined by the integral along the boundary directions of the pullback of $j^\mu$ to the boundary,
\begin{equation}\label{F def}
	{\cal F}(\delta_{\bf 1}, \delta_{\bf 2}) = \int_{\partial M} (-g)^{-1/2} n_\mu j^{\mu} d^d x, 
\end{equation}
\noindent where $n_\mu$ is the outward-pointing, unit normal to the boundary $\partial M$.
As stated in \cite{Andrade:2013wsa}, the energy density is an irrelevant operator, so we set its source to zero $h_{tx}^{(0)} = 0$.
Evaluating \eqref{F def} for the perturbations \eqref{pert Px} with boundary conditions $h_{tx}^{(0)} = 0$, we find
\begin{align}\label{F result}
\nonumber
	{\cal F}(\delta_{\bf 1}, \delta_{\bf 2}) &= - \alpha \log r \int d^d x \left[  h_{tx}^{(1)} \wedge \source^{(0)} 
	+ s^{(0)} \wedge (- i \omega \source^{(0)} ) \right] \\
	&+ \int d^d x [ h_{tx}^{(1)} \wedge ( 2 a^{(2)}_x - 2 h_{tx}^{(2)} + \tilde h_{tx} ^{(2)} ) + s^{(0)} \wedge ( 4 s_{(2)} - \tilde s^{(2)} ) ], 
\end{align}
\noindent where the wedge denotes anti-symmetrization 
\begin{equation}
	a \wedge b = a_{\bf 1} b_{\bf 2} - a_{\bf 2} b_{\bf 1}.
\end{equation}
Here we have used the asymptotic expansions \eqref{uv series1}-\eqref{uv series3} to evaluate the near-boundary limit. 
Recall that we are identifying $h_{tx}^{(1)}$ and $s^{(0)}$ with the sources of
the momentum and scalar operators, respectively. Then, from the finite piece in \eqref{F result}, we can read off the dual vevs
as
\begin{align}
 	{\cal P}_x &= 2 a^{(2)}_x - 2 h_{tx}^{(2)} + \tilde h_{tx} ^{(2)} , \\
 	{\cal O} &=  4 s_{(2)} - \tilde s^{(2)} .
 \end{align} 
Furthermore, in order to ensure that the dynamics of the system are well-defined, we must choose boundary conditions 
which make the flux vanish. The divergent piece of \eqref{F result} can be cancelled by setting the gauge invariant source $\source^{(0)}$
to zero. However, this does not suffice to eliminate the finite contribution, which requires us to impose the gauge-dependent conditions
$s^{(0)} =0$ and $h^{(1)}_{tx} =0$. A closely related behaviour was encountered in the pure Lifshitz case considered in 
\cite{Andrade:2013wsa}. One may in principle worry that imposing these two conditions separately may overdetermine the system 
of differential equations which controls the fluctuations. However, this is not the case, which becomes clear if we work in terms of gauge-invariant variables.

\bibliographystyle{JHEP-2}
\bibliography{lif_axions}

\end{document}